# Exploring the relevance of ORCID as a source of study of data sharing activities at the individual-level: a methodological discussion

Sixto-Costoya, Andrea[1,2]; Robinson-Garcia, Nicolas[3]; van Leeuwen, Thed[4]; Costas, Rodrigo[4,5]

[1,2]Department of History of Science and Information Science, School of Medicine, University of Valencia, Valencia, Spain; UISYS Research Unit (CSIC – University of Valencia), Valencia, Spain; [3]Delft Institute of Applied Mathematics, Delft University of Technology, Delft, Netherlands; [4]Centre for Science and Technology Studies (CWTS), Leiden University, Leiden, The Netherlands; [5]DST-NRF Centre of Excellence in Scientometrics and Science, Technology and Innovation Policy, Stellenbosch University, Stellenbosch, South Africa

ORCID:

Andrea Sixto-Costoya: 0000-0001-9162-8992

Nicolas Robinson-Garcia: 0000-0002-0585-7359

Thed van Leeuwen: 0000-0001-7238-6289

Rodrigo Costas: 0000-0002-7465-6462

Email of corresponding author:

andrea.sixto@uv.es

**Abstract**

ORCID is a scientific infrastructure created to solve the problem of author name ambiguity. Over the years ORCID has also become a useful source for studying academic activities reported by researchers. Our objective in this research was to use ORCID to analyze one of these research activities: the publication of datasets. We illustrate how the identification of datasets that shared in researchers' ORCID profiles enables the study of the characteristics of the researchers who have produced them. To explore the relevance of ORCID to study data sharing practices we obtained all ORCID profiles reporting at least one dataset in their "works" list, together with information related to the individual researchers producing the datasets. The retrieved data was organized and analyzed in a SQL database hosted at CWTS. Our results indicate that DataCite is by far the most important data source for providing information about datasets recorded in ORCID. There is also a substantial overlap between DataCite records with other repositories (Figshare, Dryad, and Zenodo). The analysis of the distribution of researchers producing datasets shows that the top six countries with more data producers, also have a relatively higher percentage of people who have produced datasets out of total researchers with datasets than researchers in the total ORCID. By disciplines, researchers that belong to the areas of Natural Sciences and Medicine and Life Sciences are those with the largest amount of reported datasets. Finally, we observed that researchers who have started their PhD around 2015 published their first dataset earlier that those researchers that started their PhD before. The work concludes with some reflections of the possibilities of ORCID as a relevant source for research on data sharing practices.

**Keywords**

**Introduction**

*Data sharing as a scholarly practice*

Most of the arguments supporting data sharing activities are articulated through the open research data movement, which has supported the principles and ideals of open science and data sharing across multiple scientific communities with the publication of international declarations and the support of publishers and funding agencies (Giofrè et al., 2017; Max-Planck-Gesellschaft, 2003). The main rationale behind data sharing is that it contributes to saving time, money, and efforts, and allows researchers to validate their results through the reanalysis of the shared datasets (Popkin, 2019). However, despite the increasing availability and accessibility of research data, there are different levels of data openness as well as disparate forms of recognition across countries and disciplines (Borgman, 2012; Tenopir et al., 2011). This is partially due to the lack of recognition that data production has had in the reward system of science as compared with other classical research outputs such as journal articles or books. To solve this issue, different initiatives have been launched in the past years to develop standards that aim at reducing barriers, ensuring the long-term preservation of datasets and guaranteeing authorship recognition (Womack, 2015; Make Data Count, n.d.).

There are four main different ways in which research data is shared in the scientific context: 1) by adding a dataset as supplementary material to a publication, 2) by sharing it upon request to other researchers or making it available in a personal or other website[1], 3) by uploading the dataset to a data repository, or 4) by publishing on a data journal (Federer et al., 2018; Costas et al., 2013). According to some studies, uploading datasets to data repositories is, in terms of preservation, openness and authorship recognition, the most appropriate way to share data (Hernández-Pérez & García-Moreno, 2013; Kim & Burns, 2016; Mannheimer et al., 2019). Currently, there are three types of data repositories: *multidisciplinary*, *disciplinary* or thematic, and *institutional* repositories (Pampel et al., 2013).

*ORCID as a data source to capture individual scholarly activites*

The study of academic activities of individual researchers is one of the most relevant research areas in scientometrics (Abramo, D'Angelo, & Solazzi, 2011; Costas, van Leeuwen, & Bordons, 2010; Wildgaard, Schneider, & Larsen, 2014). The focused analysis of individual researchers' activities has particular relevance for the better understanding of data sharing activities (Mongeon et al, 2017). In order to properly studying the activities of individual researchers, researcher name disambiguation is a long-standing challenge in the field of scientometrics (Smalheiser & Torvik, 2009, Caron & van Eck, 2014), finding algorithmic solutions in the most recent years (Tekles & Bornmann, 2020), although all of them have some limitations. A more fundamental solution to the problem of name ambiguity is represented by the Open Researcher & Contributor ID (ORCID), which was launched in 2012 with the aim of solving this issue. Given the challenges related to identifying individual researchers' outputs, the ORCID identifier was designed to offer the scientific community a unique registry to manage their records and information, either manually or by connecting automatically with other data sources (Brown et al., 2016; Haak et al., 2012). The purpose behind ORCID is to integrate multiple research workflows in it, so that information regarding different research outputs can easily be linked. Although journal articles and conference proceedings are among the most common outputs of scholarly activities, researchers registered in ORCID can also include more output types, for instance datasets (Fenner et al., 2015; Jefferies et al., 2019). Here we refer to datasets as all scientific data generated during the research process before being processed, raw material (spreadsheets, images, videos, etc.) that has not yet been transformed or analyzed (European Commission, 2016).

Our interest in ORCID originates from the following. ORCID is a comprehensive source of scholarly activities reported by researchers. Although sometimes limited in its uptake across scholars (Boudry & Durand-Barthez, 2020; Choraś & Jaroszewska-Choraś, 2020), it represents the only source that unequivocally relates individual scholars with their research outputs. In this paper, we focus on ORCID as a scientometric relevant *data source* to analyse data sharing practices at the individual level, which is a perspective that to the best of our knowledge hasn't been attempted before. An element that reinforces the relevance of ORCID is the fact that ORCID interlinks some of the most important data repositories to connect datasets with their researchers using their ORCID identifiers (Hernández-Pérez & García-Moreno, 2013; Kim & Burns, 2016; Mannheimer et al., 2019). The metadata registered in ORCID can be considered as being *relevant*, meaning that what is reported in ORCID by the researchers (e.g., datasets, research outputs, funding, affiliations, etc.) is likely to be correct since researchers themselves manage their own data and deliberatively include it in their public profiles. Therefore, we can assume that the act of registering datasets as research outputs in ORCID is a relatively *active form*[2] of recognizing and claiming ownership of their research outputs.

---

[1] This is the most informal form of data sharing, also the most difficult to track and measure.

Second, the presence of datasets in ORCID puts the focus on the individual researchers, who then become the key players on the promotion and recognition of data sharing practices, in contrast to other approaches that focus on institutions or repositories (Robinson-Garcia et al, 2017; Dudek, Mongeon, & Bergmans, 2019). Finally, studying data sharing practices[3] as registered in ORCID offers the opportunity to connect some of the scholars' personal traits (e.g., international and/or institutional scientific mobility, age, gender, academic status, etc.) with their outputs. This allows us to identify which institutions are adopting data sharing practices (Gómez et al., 2019), something that has not been feasible via other sources (e.g., Robinson-Garcia et al., 2016, 2017).

*Objectives of the study*

In this study, we analyse researchers' dataset outputs as reported in the ORCID platform. The specific objectives of the study are the following:

1. Quantify the number of datasets reported by individual researchers in ORCID.

2. Characterize them by the data repositories where they are registered.

---

[2] To some extent we assume (of course with some limitations) that researchers: 1) first, connect their ORCID to DataCite records (which can be seen as minimal "mindful" act towards the incorporation of datasets in their ORCID profiles); 2) second, allow DataCite (and other trusted data repositories) to automatically update their public profiles, such updates typically require a basic acknowledgement by the user (e.g. authoring trusted partners, notification e-mails, approvals by the users, etc. – see https://support.orcid.org/hc/en-us/articles/360006973133-Add-works-to-your-ORCID-record), thus being different from more automatic and algorithm updates by platforms such Google Scholar or ResearchGate; and 3) arguably researchers review once in a while their ORCID profiles in order to correct them or change the information, this could be the chance for many researchers to incorporate (or not) their dataset records.

[3] We argue that studying data sharing practices as registered in ORCID represent a relatively more "active" perspective, in which researchers registered in ORCID facilitate to some extent the identification of these practices by recording them in their profiles. This may be seen as different from more "passive" types of engagements, in which researchers may be tracked to datasets records (e.g., via records in DataCite - see Mongeon et al, 2017) but they are not necessarily including them in their CVs, research profiles or simply not linking them to their ORCID ids. Thus, we argue that in this study we are focusing on relatively "mindful" forms of data sharing activities from an individual point of view, in which researchers with a dataset recorded in ORCID have minimally provided and allowed (either mechanically – via automatic updates from trusted data repositories, or more manually) datasets to be updated and recorded in their profiles.

3. Study the breakdown of data sharing by the disciplines and countries of the individuals in ORCID.

The three objectives above are meant to provide results and empirical evidence to support a methodological discussion on the potential of ORCID as a source to capture and monitor data sharing activities from an individual point of view. After years talking about the advantages and disadvantages of sharing data and its characteristics, assuming that it is a useful practice between disciplines and researchers, the results of this study can help us to understand where this practice really goes from as reflected by the ORCID profile. That is, how researchers have adopted this practice and how much they show it in their profiles or academic profiles. In addition, it is useful to study how wide the differences in data sharing practices are among disciplines, countries, or years of research experience of the researchers.

**Methods and data**

In March 2020, we obtained from Figshare the ORCID public file. The ORCID public file contains public information associated with each user's ORCID record up to the year 2019 (Blackburn et al., 2019). The retrieved data was parsed and organized into a SQL database hosted at CWTS. A total of 7,182,036 records corresponding to distinct ORCID profiles were retrieved.

Two types of information were extracted from ORCID:

1) all ORCID profiles reporting at least one dataset in their list of "works", and

2) information related with the individual researcher producing the dataset.

In the following section we describe the data collection process as well as the information gathered for each record.

*Data retrieval and identification of sources*

Datasets were identified in ORCID by querying the field document type in the metadata of related to the "Works" of each ORCID profile (from now onwards we will use researchers when referring to those individuals who have an ORCID profile publicly available). We specifically searched for the label 'datasets' in the 'work type' field in order to unequivocally identify works records related to datasets. A total of 80,555 records (datasets) linked to 12,686 ORCID profiles were retrieved. Figure 1 illustrates how the information related with document type are showcased in ORCID.

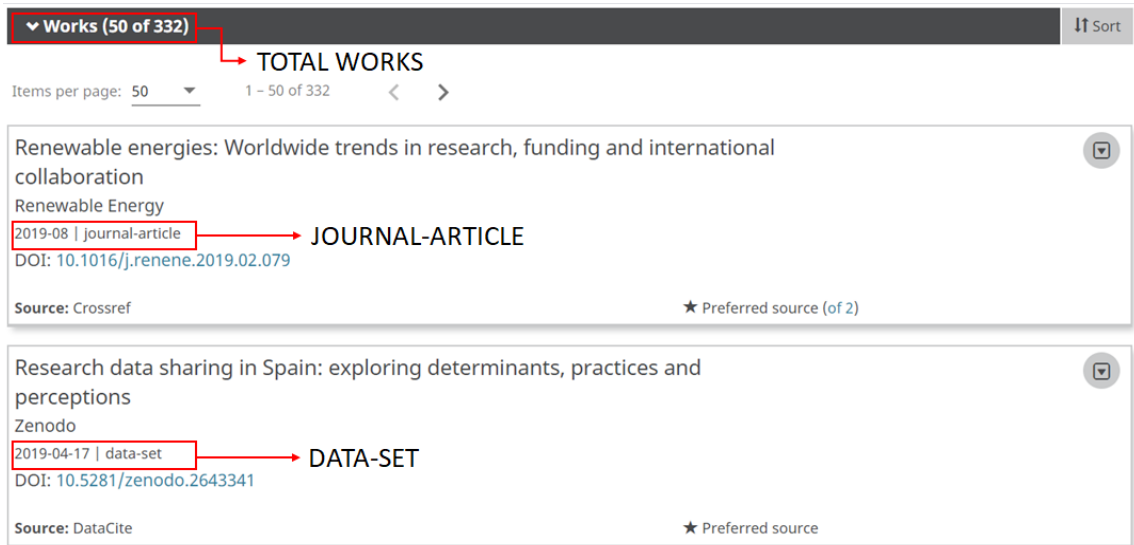

Fig. 1 Example of an ORCID profile including datasets in the scientist's output list.

For each dataset record, we extracted the source from which the information was obtained. The purpose behind this step is to better comprehend how metadata on data sharing is fed into ORCID.

In this study, we focus on four data resources: three multidisciplinary data repositories (Zenodo, Dryad and Figshare); and DataCite, a portal which offers metadata on datasets from different data repositories. We chose the three repositories due to their multidisciplinary nature and the fact that they can be easily identified through ORCID, since the DOIs from these repositories contain text strings that allows for their straightforward identification (e.g., 10.5281/zenodo.1206163 or 10.5061/dryad.9c50s/20). It is important to highlight that currently the ORCID public does not offer direct ways to identify the source from which datasets come from. We combine this information with the field source included in ORCID's metadata.

In the case of DataCite, we consider it due to its relevance as a DOI provider for datasets and to the fact that it automatically feeds into ORCID. Consequently, we consider that all data deposited in the repositories described above could or should have DataCite as a Source in ORCID (Fenner et al., 2015). With this information, we will be able to inform on 1) the number of datasets deposited in each of the three repositories, 2) the source from which ORCID obtains that information and 3) the role of DataCite as a hub in the recording and reporting of data outputs.

Figure 2 illustrates how this information was obtained. There are three possible combinations: a) identifying a dataset deposited in one of the selected repositories but being fed to ORCID via DataCite, b) a dataset deposited in one of the selected repositories which feeds from other source, and c) a dataset deposited from one of the selected repositories, but which metadata comes from a third-party. Once we applied these three types of combinations on the total datasets in ORCID (80,555) we obtained a total of 41,667 datasets belonging to 4,284 different researchers.

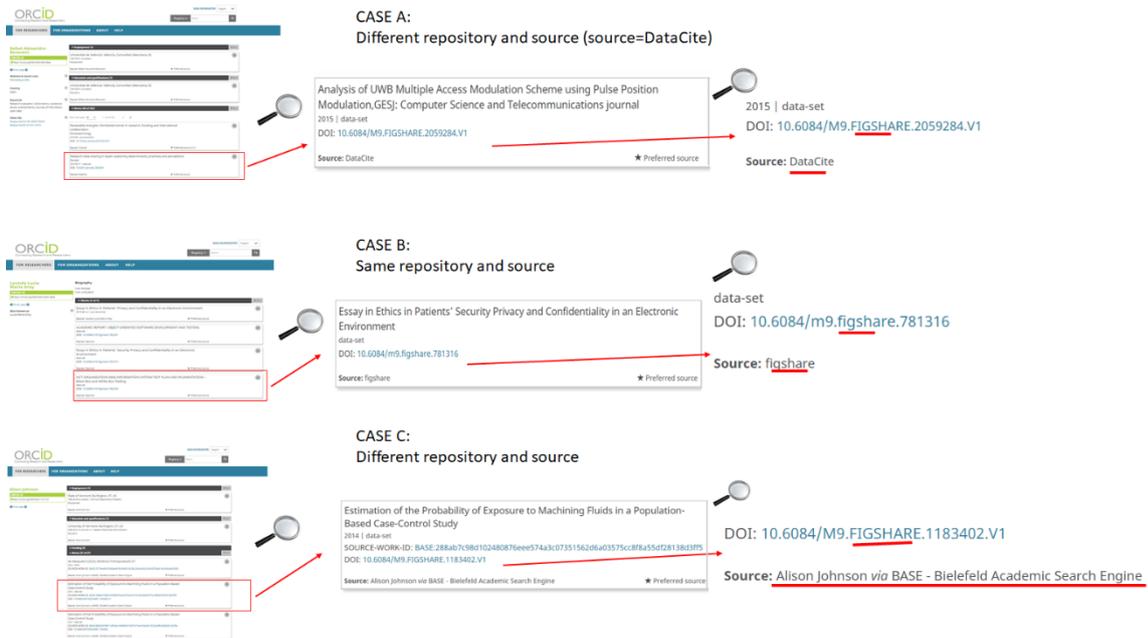

Fig. 2 Array of possible combinations on the obtention of source information of datasets in ORCID

*Extraction of individual-level variables*

For each profile identified in ORCID with datasets included as outputs, we gathered information related to the country (or countries) with which scientists were affiliated, their scientific discipline (as per linkage with the Web of Science (WoS) database), and scientific age - as defined by the date of their PhD as recorded in ORCID. In contrast to other scientometric approaches to estimate the age of researchers, like the year of first publication (Nane, Larivière, & Costas, 2017), the availability of more specific individual-level information in ORCID (e.g., year of PhD, years of education and qualifications, etc.), makes the use of such information more straightforward. In this paper we have opted to use the age of PhD as a reasonable career milestone in the academic career of researchers.

We identified the country to which authors were affiliated from the ORCID metadata, more precisely either through the employment or education and qualifications fields. We considered their most recent country/countries (could occur that one researcher is affiliated to more than one country). After obtaining the information from these two fields, we merged them in the same table. We considered both, researchers with datasets in the three repositories and DataCite (a total of 4,284 ORCID profiles) and researchers in the overall ORCID. We ended up with a total of 13,359 affiliations to 132 distinct countries in the case of the 4,284 ORCID profiles and 5,401,218 affiliations references to 247 distinct countries in the overall ORCID database.

Regarding information related to the scientific field of scientists, neither ORCID nor any of the selected data repositories include disciplinary information. To overcome this limitation, we identified their research field by matching their journal article records available in ORCID with the WoS database. From the 4,284 researchers identified with at least one dataset, a total of 3,542 had at least one publication in the Web of Science. Journal articles were matched via their DOI. A total of 88,974 journal articles were matched for the set of identified researchers with at least one dataset. Matched articles were linked to WoS subject categories. The WoS subject categories were merged following the Dutch NOWT classification available at CWTS [4]. We attributed each researcher to the different categories in a fractionalized fashion based on the distribution of their ORCID outputs across NOWT categories. We assigned researchers to the discipline in which most of their papers were classified.

Finally, we computed the academic age of scientists by considering the PhD start date located in the field education and qualifications of ORCID. More specifically, we were interested in the PhD starting date of the identified researchers and used it as a fixed point of reference in the academic career of the researchers in order to assess when they published their first dataset or journal article. We used a 5-year window of dataset publication to allow ORCID researchers to publish datasets or journal articles. Out of the 4,284 researchers (ORCID profiles) with at least one dataset in the three repositories and DataCite, 45% stated their PhD start in ORCID. As a comparative set, we also identified all the profiles in the overall ORCID field that disclosed their PhD start year, finding a total of 686,204 ORCID profiles. It is important to highlight that the information about the academic status of the researchers is presented as free text in ORCID, meaning that in order to identify the "PhD" status of the researchers, it was necessary to perform a substantial cleaning and textual processing of the information. To keep consistency, we only considered academic status indicators written in English.

We observed that most datasets were concentrated within the cohort of researchers starting their PhD between 2010 and 2015. Before this period of time, we observed that from 1970 (the first year that one researcher publishes at least one dataset in the next 5 years after starting their PhD) to 2009, the number of scientists reporting at least a dataset is very scarce (11.9% of a total of 310 researchers). Therefore, we decided to consider the period between 2010-2015 for a more detailed analysis, in which we find the larger set of individuals with at least one shared dataset (69.4%). The rest (11.6%) belongs to researchers that started their PhD from 2016 onwards and do not fit in the 5-year window.

In Table I we describe the datasets mentioned in this section in order to help the reader to understand the sequence of numbers.

Figure 3 illustrates where his information is located in ORCID for the three discussed metadata fields.

---

[4] The NOWT classification is a grouping of WoS Journal Subject Categories (JSC), whereby each JSC is attached to one level each time (without overlaps). The NOWT classification was designed in the light of the Dutch Observatory of Science & Technology, and functioned as that instrument's field classification system for over 30 years. The system contains various levels of aggregation, whereby the lowest level of aggregation consists of 37 scientific disciplines, and the highest level of aggregation consists of 7 larger domains of scholarly activity. In this study, that highest level is used.

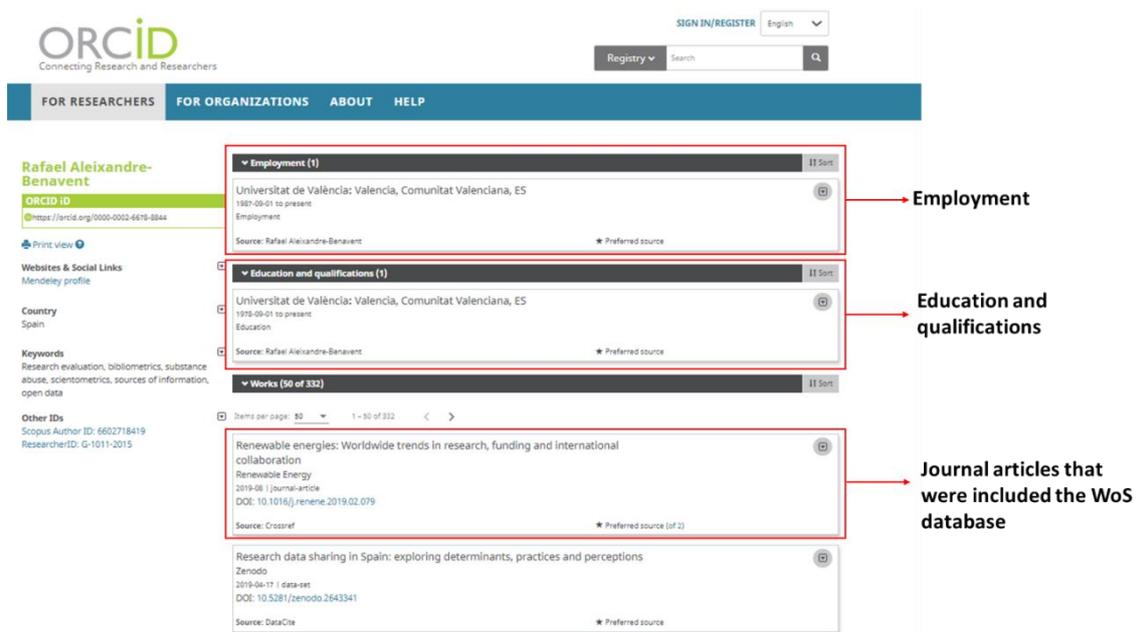

Fig. 3 Fields used in ORCID to identify scientists' affiliation country, discipline and academic age.

Table I. Description of each data mentioned in the "methods" section

| Datasets | Description of the datasets |
|---|---|
| 7,182,036 | Total profiles in ORCID |
| 12,686 | Total profiles in ORCID with at least one dataset |
| 80,555 | Total distinct datasets in ORCID |
| 4,284 | Total profiles in ORCID with at least one dataset indexed to Zenodo, Dryad, Figshare or DataCite |
| 41,667 | Total distinct datasets in ORCID related to Zenodo, Dryad, Figshare or DataCite |
| 13,359 | Total set of researcher-affiliation linkages from the 4,284 profiles with at least one dataset related to Zenodo, Dryad, Figshare or DataCite, referring to a total of 132 different countries |
| 5,401,218 | Total set of researcher-affiliation linkages. These affiliations refer to a total of 247 different countries in the total ORCID profiles |
| 3,542 | ORCID profiles with at least one journal article in WoS of the 4,284 profiles with at least one dataset related to Zenodo, Dryad, Figshare and DataCite |
| 88,974 | Total journal articles that were matched in WoS for the 4,284 profiles with at least one dataset related to Zenodo, Dryad, Figshare or DataCite. Of them, 3,542 have one article or more in WoS. |

**Results**

We now report our findings by focusing on the total number of datasets in Zenodo, Dryad, Figshare and DataCite. We took into account the following three scenarios:

1) the repository is Zenodo, Dryad or Figshare and the source is DataCite,
2) the repository is Zenodo, Dryad or Figshare but the source is not DataCite, and
3) the source is DataCite but the repository is none of those three.

Considering these criteria, we found a total of 41,667 datasets connected to a total of 4,284 different researchers.

Therefore, when onwards the expression "researchers with datasets" or similar appears in the following sections, we are referring to this group of people with their corresponding datasets. We structure this section in the following way. First, we report differences by repository and overlap with DataCite. Second, we focus on the information related with the individual researcher producing the dataset.

*Repositories and overlap with DataCite*

DataCite is by far the most important data source in providing information about datasets as recorded in ORCID. However, there is an important overlap between DataCite as source with the other repositories (Figshare, Dryad and Zenodo). That means that DataCite works as a general aggregator wide collector of datasets from other resources with a noticeable influence on ORCID workflows (Fig. 4).

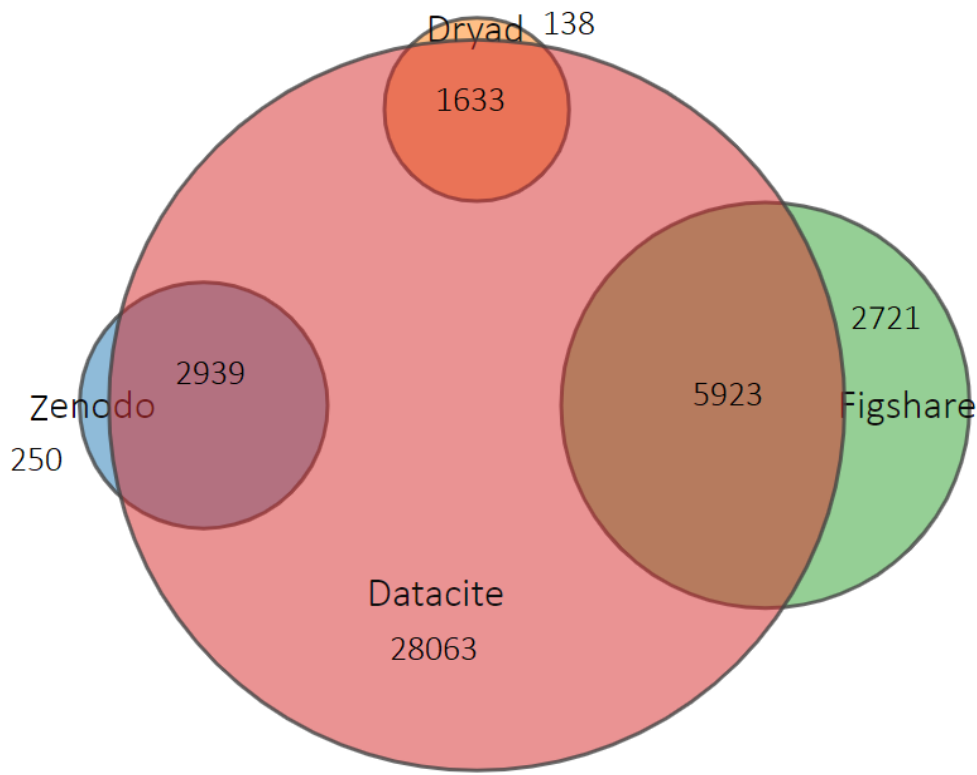

Fig. 4 Number of data sets in Zenodo, Dryad or Figshare whose source is either DataCite or other and the total number of datasets whose source is DataCite

*Analysis by country*

Researchers in our sample are (or have been) affiliated to institutions belonging to a total of 132 different countries. Figure 5 shows the top 15 countries with researchers that have reported datasets in ORCID. We observe that in the case of the top 6 countries, up to Canada, the trend is to have a higher percentage of people who have datasets with respect to the total of people with datasets than researchers in the total ORCID. We observe the opposite pattern in the rest of countries included in Figure 5.

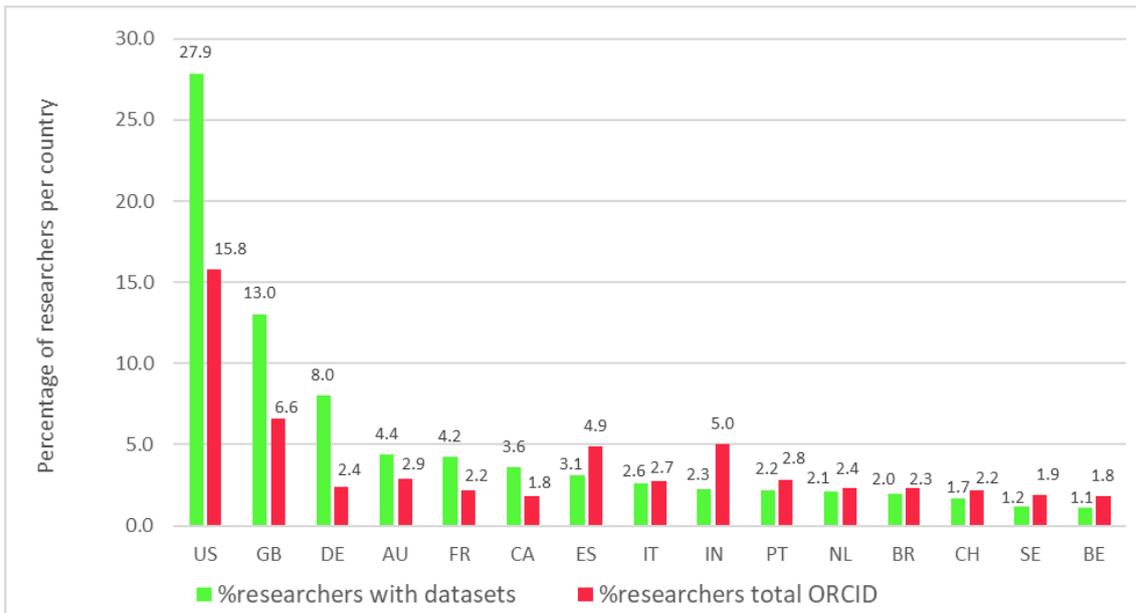

Fig. 5 Average of researchers with datasets per country in relation to total number of researchers with datasets and the total number of researchers in ORCID. Countries are referred using their corresponding ISO codes.

*Analysis by discipline*

In Figure 6 we can observe the percentage of researchers by each NOWT category. These percentages show us to which scientific area the journal articles published by researchers with ORCID datasets belong. With these results we have an approximation of the areas where more datasets are shared in ORCID.

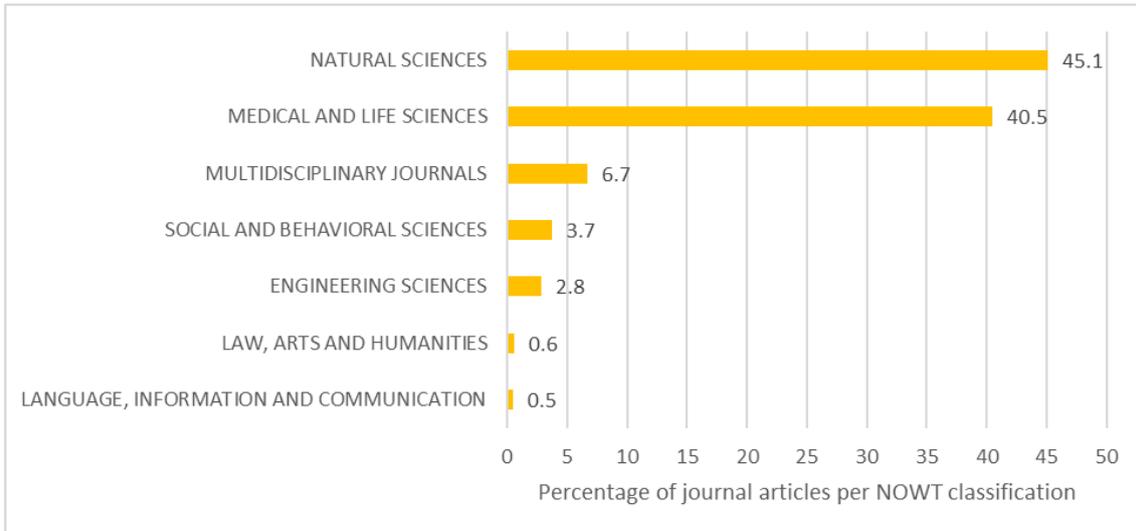

Fig. 6 Distribution of scientists by discipline according to the NOWT classification

*Academic age of scientists sharing datasets*

Figure 7 shows that researcher who started their PhD more recently, published their dataset in a shorter period of time, particularly if we compare to the average time it took them to publish their first papers, that exhibits a more constant average pattern.

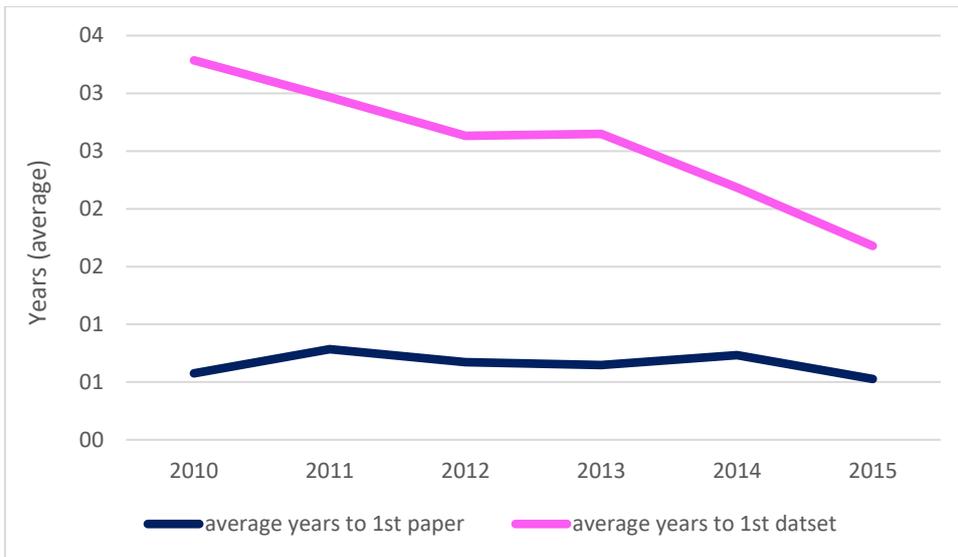

Fig. 7 Average of years that researchers take to publish their first dataset and journal article for the 2010-2015 period.

**Discussion and conclusions**

This paper presents a discussion about the methodological possibilities of using information recorded in ORCID as a data source to study global data sharing practices from an individual point of view. Since we focus on one main source to study data sharing practices, it is also important to highlight the limitations of ORCID as source itself (Youtie, Carley, Porter, & Shapira, 2017). Among these limitations we need to mention that ORCID is not necessarily globally adopted platform, among all researchers and across all disciplines and countries (Boudry & Durand-Barthez, 2020). Moreover, there may be some concern about data protection and privacy issues, which could cause some refusal to create an ORCID profile (Choraś & Jaroszewska-Choraś, 2020). However, despite these limitations, recent studies observe a remarkable growth of ORCID between countries and organizations, which gives the opportunity to analyse various aspects related to training, origin or the different types of publications (original articles, datasets, conference proceedings…) belonging to researchers (Arunachalam & Madhan, 2016; Gomez, Herman, & Parigi, 2020).

In this study, we focus on datasets recorded in ORCID from four major data resources: Zenodo, Dryad, Figshare and DataCite. First, we must note that these four sources represent two general types of information sources which differ conceptually. One type is represented by DataCite, which allows to find and download datasets but is not a data repository itself, but rather a DOI registering organization for datasets. The second type is represented by Zenodo, Dryad and Figshare, repositories which allow researchers to upload and preserve their datasets but do not collect information from other repositories (Enis, 2013; He & Han, 2017; Neumann & Brase, 2014; Sicilia et al., 2017).

Another important limitation is that we look at the overlap across the selected sources (DataCite sometimes captures information that has already been captured by other resources such as Zenodo or other repositories and vice versa) that could be due to a conflict between the manual and automatic modalities or other technical problems. For example, a researcher manually adds a record and DataCite adds the same record later creating a duplicate. In addition, the manual option could cause problems when researchers fill in the gaps to enter a new record. In this sense, we identified that sometimes researchers introduce erroneous information that can cause misunderstandings about, for example, what is the source of the aggregated information. Moreover, we found that it is quite difficult to identify the name of the repositories, which we found thanks to the DOI string but only for those data repositories with this feature (i.e., Zenodo, Figshare and Dryad).

However, against this background of ORCID-related limitations, we argue that we are interested in studying data sharing practices that have some degree of active attitude by individual researchers. Thus, data sharing activities that are captured by repositories, or that come as a result of publishers' and funders' policies but without an active acknowledgement by individual researchers as genuine outputs worth of including in their research profiles, fall out of the interest of this paper. As a matter of fact, for this type of analysis, only tools like ORCID as well as other types of profile and individual-level oriented research platforms (e.g. Curriculum Lattes in Brazil, Curriculum Vitae Normalizado in Spain, researcher's individual websites, ResearchGate or Academia.edu) could be seen as relevant sources that could provide this notion.

The results of our study confirm that DataCite is a central tool for locating and identifying datasets. DataCite has already been documented in previous studies (Mongeon et al., 2017; Robinson-Garcia et al., 2017), thus reinforcing the role of DataCite as a pivotal resource to study data sharing practices. DataCite is also a very interesting resource from a user perspective, since it unifies the information of multiple data repositories in one single place. That means that DataCite offers the user the opportunity of finding datasets from different data repositories while the other three ones analyzed (Figshare, Zenodo and Dryad) do not record information of data from other resources.

Regarding differences among researchers' countries, we consider for each researcher his or her most recent country/countries listed in his/her profile. We are aware that other strategies could have been considered, such as trying to relate the datasets to the country of the researcher at the time of publication. However, taking into account the characteristics of ORCID as a tool, where correctly tracing this relationship is very difficult, we consider the option of selecting the most recent country to be more appropriate.

As for our results in relation to the country analysis that we finally conducted, they reflect the uptake of both ORCID and data sharing practices in different national and disciplinary settings. In the case of ORCID uptake, the top 15 countries to which researchers in our sample belong to, have actively promoted the implementation of ORCID. Hence, in the UK we find specific initiatives such as the one made by the UK Data Service about a project with the British Library to enhance the use of ORCID in the Academia (Imperial College of London, n.d.; Meadows, 2015; UK Data Service, n.d.). Brazil has launched initiatives such as the Brazil Consortium, a collaboration between CAPES (a funding agency under the Ministry of Education) and ORCID launched in 2018. Or in the case of Spain, the initiative of the Spanish Foundation for Science and Technology (FECYT) that announced in 2019 an agreement to join the ORCID community (FECYT, 2019; Heredia, 2018). In the same line, the Canadian Research Knowledge Network also promoted with other members of the Canadian academic community a consortium with ORCID, following the experience of other countries such as the UK or Australia (Canadian Research Knowledge Network, 2020.

In the case of data sharing, we observe similar initiatives in the top countries regarding the number of datasets produced by the researchers registered in ORCID. In the US, several initiatives coexist in different public and private institutions. For instance, the National Institutes of Health (NIH) and the National Science Foundation (NSF) (National Institutes of Health, 2015; Womack, 2015) require data sharing plans to ensure transparency and reusability of funded research. In Canada, the Canadian Institutes of Health Research have a well-established policy of data sharing that implies the publication and deposit of data in open data repositories after the results have been published (Canadian Institutes of Health Research, 2015). Another example is the Australian National Data Service (ANDS) that has been encouraging data sharing among researchers since 2008 (Guru et al., 2013). Moreover, the high presence of European countries is probably explained due to the strong commitment of the EU in data sharing, that could be observed through the data sharing policy of the Horizon 2020 programme (European Commission, 2016).

Our results show relevant differences between disciplines in the publication of datasets. The disciplines with more presence in the ORCID profiles of researchers with datasets are related to Natural Sciences and Medical and Life Sciences. The case of Natural and Life Sciences was studied in several other previous studies concluding that data sharing in the group of disciplines belonging to this area is higher and better assumed than in other areas (Tenopir et al., 2011, 2015). For example, disciplines such as Meteorology and Physics data sharing practices have been common for decades, as well as in other areas like genetics and genomics large scale datasets were crucial to enable new and important discoveries in those disciplines. In the case of Ecology, the exchange of data between researchers around the world allowed this field to obtain results which would have been very difficult to obtain without the expansion of data sharing practices(Nature Communications, 2018; Sieber, 2015; Berghmans et al., 2017). A good reasoning for the situation of these disciplines with respect to data sharing was given by (Sieber, 2015), who explains that in very expensive sciences such as astronomy, oceanology, and space exploration the researchers decided to share not only data but also equipment as telescopes to advance more quickly. Moreover, in the case of this disciplines, to share data is also necessary because the amount of data to be analyzed is simply too much to get the job done by academics alone (Hoeppe, 2014; Pepe, Goodman, Muench, Crosas, & Erdmann, 2014).

Regarding the analysis of the relationship between the academic experience in the research career regarding the publication of datasets, our study shows that there is a relationship between the year of beginning of the PhD and the year of publication of the first set of data: the more recently the researchers started their PhD, the more possibilities there are to publish the first dataset in the following 5 years, which reinforces the idea that new generations of researchers are embracing more actively data sharing activities, compared to previous generations. This observation is in line with other previous studies that reported the belief by early career researchers that the data sharing could help them to boost their scientific career (Gewin, 2016). Moreover, the scientific literature indicates that the older an article is the less likely to have data availability. For example, in the study done by Vines et al., (2014), they found a clear decrease of the dataset availability when the articles were getting older and this was partially explained by technological problems of storage and retrieving. Popkin (2019) claims that there is a generational change before and after the "digital age" in the sense that before sharing data was generally limited to sending it on request but now, that data can be shared instantly if there is an Internet connection, there is an increase in data exchange due to these facilities, which also indicates how the technologies linked to the use of Internet have influenced data sharing practices; including also the development of infrastructures for data repositories, which contributed to enable and facilitate the sharing of data by researchers. Furthermore, it is important to highlight that for approximately ten years numerous institutions, governments, universities and other research centres, as well as journals and publishers, have echoed the advantages of data sharing and promoted it in different ways so it is obvious that we are living, in the light of open science, a moment of debate and it is also not surprising that this movement is especially affecting the new generations of researchers (Alsheikh-Ali, Qureshi, Al-Mallah, & Ioannidis, 2011; Popkin, 2019; Sholler, Ram, Boettiger, & Katz, 2019).

**Final remarks**

From a methodological point of view, ORCID can be seen as a relevant tool for the study of the inclusion of dataset outputs in the research profiles, contributing to some extent to align with the FAIR principles. First, an important conceptual advantage of ORCID as a data source to study scholarly practices, is that it somehow also captures the individual perception of these activities. Thus, when focusing on data sharing practices, the information recorded in ORCID profiles about datasets can be considered as providing some idea about the self-perception of datasets as first order research outputs (at a similar level as journal articles). Second, we demonstrated that ORCID itself is a useful resource to know not only the information about the researcher's outputs but also biographical information that is very useful for mapping who is doing what and where. For this reason, to the promotion and adoption of ORCID among the scientific community becomes of paramount importance, since it offers the scientific community a unique tool to study its main dynamics, not only in data sharing practices as demonstrated in this paper, but also in other aspects such as mobility or career development to name a few (Yan, Zhu & He, 2020).

## Acknowledgments

This work was supported by Ministry of Science and Innovation of Spain (BES-2016-079394) and European Social Fund and was partially funded by the South African DST-NRF Centre of Excellence in Scientometrics and Science, Technology and Innovation Policy (SciSTIP). We thank an anonymous reviewer for insightful comments and recommendations of an early version of this paper.

## Author contributions

ASC, NRG, TvL and RC contributed to the conception, design, and analysis of the study, as well as writing of the manuscript.

## Compliance with ethical standards

Conflict of interest

The authors have no conflicts of interest to declare that are relevant to the content of this article.